\documentclass[reprint,aps,tightenlines,onecolumn,notitlepage,superscriptaddress]{revtex4-1}
\usepackage{graphicx}
\usepackage{amsmath}
\usepackage{amssymb}
\usepackage{csquotes}
\MakeOuterQuote{"}
\usepackage{bm}

\begin{document}

\title{Structure and overstability of resistive modes with runaway electrons}

\author{Chang Liu}
\email{cliu@pppl.gov}
\affiliation{Princeton Plasma Physics Laboratory, Princeton, New Jersey 08540, USA}
\author{Chen Zhao}
\affiliation{Princeton Plasma Physics Laboratory, Princeton, New Jersey 08540, USA}
\author{Stephen C. Jardin}
\affiliation{Princeton Plasma Physics Laboratory, Princeton, New Jersey 08540, USA}
\author{Amitava~Bhattacharjee}
\affiliation{Princeton Plasma Physics Laboratory, Princeton, New Jersey 08540, USA}
\affiliation{Princeton University, Princeton, New Jersey 08544, USA}
\author{Dylan P. Brennan}
\affiliation{Princeton University, Princeton, New Jersey 08544, USA}
\author{Nathanial M. Ferraro}
\affiliation{Princeton Plasma Physics Laboratory, Princeton, New Jersey 08540, USA}

\begin{abstract}
	
We investigate the effects of runaway electron current on the dispersion relation of resistive magnetohydrodynamic (MHD) modes in tokamaks. We present a new theoretical model to derive the dispersion relation, which is based on the asymptotic analysis of resistive layer structure of the modes. It is found that in addition to the conventional resistive layer, a new runaway current layer can emerge whose properties depend on the ratio of the Alfv\'{e}n velocity to the runaway electron convection speed. Due to the contribution from this layer, both the tearing mode and kink mode will have a real frequency in addition to a growth rate. The derived dispersion relation has been compared with numerical results using both a simplified eigenvalue calculation and a M3D-C$^{1}$ linear simulation, and good agreement is found in both cases.

\end{abstract}

\maketitle

\section{Introduction}

In tokamak discharges, suprathermal electron populations can be generated through the so-called runaway mechanism both during quiescent operation\cite{paz-soldan_growth_2014} and disruption events\cite{eidietis_control_2012}. These high-energy runaway electrons pose a serious threat to the successful operation of large tokamaks\cite{lehnen_disruption_2011}, and remain one of the outstanding challenges for ITER\cite{martin-solis_formation_2017,boozer_pivotal_2018,breizman_physics_2019}. During a disruption, runaway electrons can be formed through the Dreicer process\cite{dreicer_electron_1959,dreicer_electron_1960,connor_relativistic_1975}, the hot-tail mechanism\cite{smith_runaway_2006} or the avalanche mechanism\cite{rosenbluth_theory_1997,liu_adjoint_2017}. The avalanche growth rate of runaway electrons will be much larger in ITER disruptions compared to current tokamak devices, given the larger pre-disruption current and the induced toroidal electric field, as well as longer wall time for electric field diffusion. It is predicted that in a typical disruption event in ITER, a significant fraction of pre-disruption current will be converted to be carried by runaway electrons with energies of MeVs\cite{boozer_theory_2015,boozer_runaway_2017}. In addition, the impurities injected to mitigate other disruption consequences such as wall forces during vertical displacement events (VDE) can even worsen the runaway electron avalanche issue through partial screening effects\cite{hesslow_effect_2017,hesslow_effect_2018,hesslow_influence_2019,breizman_physics_2019}. Uncontrolled loss of REs can lead to melting of plasma facing components\cite{reux_runaway_2015}, thus it is important to study the confinement of RE beams during disruption events.

The generation of runaway electrons is very sensitive to the local plasma temperature and the parallel electric field\cite{helander_runaway_2002}. In addition, the generated runaway electrons can only survive long enough in regions with closed field lines to give rise to avalanche generation. These lead to more peaked current profiles than predisruption\cite{eriksson_current_2004}, which alters the MHD stability of the plasma. In recent disruption experiments at DIII-D, significant MHD instabilities have been observed coincident with a large generated runaway current\cite{paz-soldan_kink_2019}. The MHD instabilities can lead to deconfinement of both runaway electrons and bulk plasma, with significant oscillations observed in ECE and interferometer signals\cite{pucella_overview_2019}. To understand the effect of runaway electrons on MHD instabilities, efforts have been taken on both theoretical analysis and numerical simulations\cite{helander_resistive_2007,cai_influence_2015,matsuyama_reduced_2017,bandaru_simulating_2019}. In these works, it is found that the linear growth rate of MHD instabilities is approximately the same as in a plasma without runaway electrons, but the nonlinear saturation level is different\cite{helander_resistive_2007}. However, the perturbed current profile is much more peaked near the rational surface\cite{matsuyama_reduced_2017}. These studies are conducted using a fluid description of runaway electrons and a current coupling scheme into the MHD equations, which can save computation time compared to electromagnetic kinetic simulation\cite{liu_conservative_2018}. Recently, this method has been implemented in the M3D-C$^{1}$ code, which is a fully 3D nonlinear MHD code based on finite element method\cite{jardin_high-order_2007,ferraro_calculations_2009}. Both $m>1$ ($m$ is the poloidal mode number) tearing modes and resistive kink modes in the presence of runaway electrons have been studied using M3D-C$^{1}$, and several new interesting phenomena have been found in simulations\cite{zhao_simulation_2020}.

One of the newly found phenomena observed in our M3D-C1 numerical simulation of MHD tearing instabilities in the presence of runaway electrons is that the mode has a real frequency (overstability) driven by the runaways and is propagating in the plasma frame.  When real frequencies are driven in resistive MHD instabilities, it is critically important to understand the physics driving them and their effect on the mode, especially when studying experimental phenomena. Overstability can be highly impactful both to the stability of resistive MHD modes and the physics of their interaction with the plasma and resistive wall. The real frequency of the mode has been studied as an effect of plasma pressure gradient(\cite{coppi_resistive_1966,glasser_resistive_1975,glasser_resistive_1976,iacono_stability_1994}), two-fluid effect\cite{coppi_influence_1964,coppi_currentdriven_1965,biskamp_drift-tearing_1978,finn_driftresistive_1983}, polarization drifts\cite{waelbroeck_rotation_1997}, energetic ion interaction\cite{hegna_suppression_1989,halfmoon_model_2017}, effects of favorable average curvature\cite{akcay_nonlinear_2020}, and coupling with other modes\cite{fitzpatrick_stability_1994}. It has been used to explain the Maxwell torque and plasma rotation as mode locking happens\cite{finn_error_2015,finn_real_2019}. With multiple effects simultaneously present, a single frequency is generally found for the mode depending on the combined influences.  Under different conditions, each individual driving mechanism can be particularly important to the stability of the mode.  

In this paper, we present a new mechanism causing the mode rotation that can explain the real frequency observed, which is the attributable to the convection of runaway electrons. This work is based on the theoretical derivations in Ref. \cite{helander_resistive_2007}, where asymptotic analysis has been utilized to study reduced MHD equations\cite{furth_tearing_1973} plus a runaway electron convection equation in a cylindrical plasma. In addition to the classical resistive layer theory, we find that runaway electrons can cause a secondary layer structure to form inside the resistive layer, whose width depends on the value of $v_{A}/c$ ($v_{A}$ is the Alfv\'{e}n velocity and $c$ is the speed of light) and can be much narrower than the resistive layer. In addition to the $m>1$ tearing mode, we also use the model to study the $m=1, n=1$ resistive kink mode with runaway electrons based on Ref. \cite{coppi_resistive_1976}, whose eigenfunction of perturbed flux behaves very differently than the $m>1$ tearing mode.

This paper is organized as follows. In Sec. \ref{sec:tearing} the resistive tearing mode in the presence of runaway electrons with $m>1$ is studied, and a new dispersion relation is derived which gives both the growth rate and the real frequency. In Sec. \ref{sec:kink} a similar method is applied to study the $m=1$ resistive kink mode. The results obtained in both sections are compared with numerical calculations using both a 1D eigenvalue code and the M3D-C$^{1}$ MHD simulation code, which is presented in Sec. \ref{sec:simulation}. In Sec. \ref{sec:conclusion} we summarize. 

\section{Resistive tearing mode with $m>1$}
\label{sec:tearing}
We first study a tearing mode with $m>1$ in a periodic straight cylinder with its axis in the $z$ direction. In the framework of reduced MHD, the magnetic field and the the plasma flow can be conveniently represented as
\begin{align}
    \mathbf{B}=B_T \mathbf{e}_z+\mathbf{e}_z\times \nabla\psi,\\
    \mathbf{v}=\mathbf{e}_z\times\nabla\phi,
\end{align}
where $\psi$ is the normalized magnetic flux ($\psi=A_{z}/(B_{T}a)$, $a$ is minor radius of the cylinder), $\phi$ is the normalized electric potential ($\phi=\Phi/(B_T a v_{A})$). The linearized equations of motion of the magnetic fields, the perpendicular velocity fields (represented by electric field since $\mathbf{E}_{\perp}=\mathbf{v}_{\perp}\times\mathbf{B}_{z}$), and the convection equation of runaway electrons can be written as\cite{helander_resistive_2007}
\begin{align}
\gamma \psi+ik_\parallel \phi =\eta (\hat{\nabla}^2_\perp \psi+j_{RE}),\label{eq:gammapsi}\\
\gamma \hat{\nabla}_\perp^2\phi+ik_\parallel \hat{\nabla}_\perp^2\psi =-\frac{im J_0'}{r}\psi, \label{eq:gammaphi}\\
(k_\parallel +i\gamma v_A/c)j_{RE}=\frac{m J_{RE0}'}{r} (\psi+\phi v_A/c).\label{eq:jre}
\end{align}
Here we ignore the effects of plasma pressure, given that the plasma $\beta$ is close to zero in post-disruptions. $\gamma$ is the linearized growth rate normalized to Alfv\'{e}n time $\tau_A=a/v_{A}$, $j_{RE}$ is the perturbed runaway electron current density, $J_0$ is the equilibrium current density, $J_{RE0}$ is equilibrium runaway electron current density, $\eta$ is the normalized resistivity ($\eta=\eta_\parallel/(\mu_0 v_{A}a)$), and $m$ and $n$ are the poloidal and toroidal mode numbers.  In addition,
\begin{equation}
k_\parallel=\frac{n}{R}\left(\frac{m}{nq}-1\right),
\end{equation}
\begin{equation}
\hat{\nabla}_\perp^2 A=\frac{1}{r}\frac{d}{dr}\left(r\frac{dA}{dr}\right)-\frac{m^2}{r^2}A, \quad A'=\frac{dA}{dr},
\end{equation}
where $R$ is the major radius, and $q=r B_T/(R B_P)$ is the safety factor, with $B_P=|\nabla\psi|$ being the poloidal field. Note that in Eq. (\ref{eq:jre}), the runaway electrons are assumed to have a convection velocity equal to $c$ on the opposite direction of the magnetic field line (so the runaway electron current is positive), plus the $\mathbf{E}\times\mathbf{B}$ drift, given that most runaway electrons are relativistic with $v_{\parallel}\gg v_{\perp}$.

Note that in Eqs (\ref{eq:gammapsi}-\ref{eq:jre}), the gradient and curvature drift of of runaway electrons are ignored. For high energy runaway electrons, the curvature drift can be important and comparable to the $\mathbf{E}\times\mathbf{B}$ drift. However, the magnitude of these drift terms depends on runaway electron energy, which is missing in the fluid description of runaway electrons. In addition, the effect of curvature drift can be cancelled by vertical fields from external coils to maintain the force balance\cite{hu_inward_2016}. The pressure of runaway electrons is also absent, since we are using a current coupling scheme rather than pressure coupling for runaway electrons. The effects of these terms on MHD stability will be studied in the future. 

Given that $\gamma\ll 1, \eta\ll 1$, for the solution far away from the rational surface (outer region), terms proportional to $\eta$ and $\gamma$ can be ignored. The solution $\psi_{\mathrm{out}}$ can be used to determine the boundary condition of the inner region. For the solution close to the rational surface, $k_\parallel \to 0$ and the resistive term becomes important. We introduce the coordinate variable $x=r-r_s$, where $r_s$ is the minor radius of the rational surface. The equations in the inner layer can be rewritten as
\begin{equation}
\label{eq:psipp}
\psi''+\frac{1}{r_s}\psi'-\frac{m^2}{r_s^2}\psi+j_{RE}=\frac{\gamma}{\eta}(\psi+x\xi),
\end{equation}
\begin{equation}
\label{eq:xipp}
\frac{\gamma^2}{k_c^2}\xi''=x\left(\psi''+\frac{1}{r_s}\psi'-\frac{m^2}{r_s^2}\psi\right)-a\psi-bx\psi,
\end{equation}
where $x\ll r_s$ is assumed, $\xi=ik_c \phi/\gamma$ is the plasma displacement, $k_c=n q'/(Rq)$, and \cite{militello_effects_2004}
\begin{equation}
a=\frac{m}{k_c r_s}J_0',
\end{equation}
\begin{equation}
b=\frac{1}{k_c}\left[a\left(\frac{q''n}{2q R}\right)-\frac{d}{dr}\left(\frac{m}{r}J_0'\right)\right].
\end{equation}

Note that the resistive term is only important in the layer $|x|<|\delta_1|$, where
\begin{equation}
\label{eq:delta-1}
    \delta_1=\gamma^{1/4}\eta^{1/4}k_c^{-1/2}.
\end{equation}
The last two terms on the right-side of Eq. (\ref{eq:xipp}) come from the right-side of Eq. (\ref{eq:gammaphi}), which are ignored in \cite{furth_tearing_1973} in the inner layer calculation. By including the first order contribution of these two terms, we can obtain a corrected value of the resistive tearing mode growth rate, as discussed in \cite{militello_effects_2004}.

The perturbed runaway electron current in Eq. (\ref{eq:jre}) can be rewritten as follows in the inner layer,
\begin{equation}
\label{eq:jre-layer}
j_{RE}=\frac{m J_{RE0}'}{k_c r_s}\frac{\psi+v_A\phi/c}{x+i\gamma v_A/(k_c c)}.
\end{equation}
Assuming $J_{RE0}=J_{0}$, and ignoring the terms proportional to $v_A/c$ in Eq. (\ref{eq:jre-layer}), we find that the runaway electron current term can cancel the last two terms in Eq. (\ref{eq:xipp}), leaving the equation exactly the same as the approximate equation in \cite{furth_tearing_1973}. Thus the growth rate will follow the 3/5 power law with respect to $\eta$ in \cite{furth_tearing_1973}, as discussed in \cite{helander_resistive_2007}. For $J_{RE0}\ne J_{0}$, the last two terms in Eq. (\ref{eq:xipp}) will then come only from the thermal electron current $J_0-J_{RE0}$, and the mode growth rate will be reduced by a certain factor.

We now consider the effect of finite value of $v_A/c$. Given $v_A\ll c$, the term proportional to $v_A/c$ in the numerator in Eq. (\ref{eq:jre-layer}) can be ignored. However, the second term in the denominator leads to an imaginary correction to $j$ for $x\to 0$. Define
\begin{equation}
\label{eq:delta-2}
   \delta_2=\gamma v_A/(k_c c), 
\end{equation}
we find that $|\delta_2|\ll|\delta_1|$, thus this correction is only significant in a sublayer inside the inner layer. To satisfy Eq. (\ref{eq:psipp}), we can therefore assume $\psi''=-j_{RE}$ inside this sublayer. Following the constant-$\psi$ approximation, it is found that the two terms on the right hand side of Eq. (\ref{eq:psipp}) are subdominant compared to $\psi''$, given $|\delta_{2}|\ll \gamma/\eta$. Therefore 
\begin{equation}
\label{eq:inner-matching}
\int_{-\delta_2}^{\delta_2} \psi'' dx\approx \pm i\pi \frac{m J'_{RE0}}{k_c r_s} \psi(x=0).
\end{equation}
Here we apply the Sokhotski–Plemelj theorem to calculate the complex integral and ignore the principle value, which is similar to the calculation of the imaginary term in a Landau contour integral. The sign of the integration result will be the same as the sign of the term $\gamma v_A/(k_c c)$. Note that the real part of $\psi$ will also be affected by $j_{RE}$ and behave like $\psi=-\psi_0[1+mJ'_{RE0}/(k_c r_s) x\ln |x|]$ near $x=0$, as discussed in \cite{helander_resistive_2007}. This part will also contribute to the matching condition through the last two terms on the right hand side of Eq. (\ref{eq:xipp}).

To match to the ideal MHD solution in the outer region, the solution needs to satisfy
\begin{equation}
\label{eq:outer-matching}
\int_{-\delta_1}^{\delta_1} \psi'' dx=\Delta' \psi_0,
\end{equation}
where $\Delta'=(\psi'(x=0+)-\psi'(x=0-))/\psi_0$, where $\psi'(x=0+)$ and $\psi'(x=0-)$ are taken from the solution in the outer region. Here we take the "constant-$\psi$" approximation in the layer by assuming $\psi=\psi_0$.

For the region $|\delta_2|\ll |x| \ll |\delta_1|$, the correction caused by finite value of $v_A/c$ is very small. In addition, as discussed above, the runaway electron current term can cancel the last two terms in Eq. (\ref{eq:xipp}), leaving the equation exactly the same as the approximate equation in \cite{furth_tearing_1973}. We can thus assume that the functional form of the solution of $\xi$ is little affected by the existence of the sublayer. Note that the amplitude of the solution, including both real and complex parts, will be affected by the sublayer as shown below.

Introduce $z=x/\delta_1$, $\chi=\delta_1\xi/\psi_0$, then in the region $|z|\gg |\delta_2|/|\delta_1|$, the equation can be simplified as
\begin{equation}
\chi''-z^2\chi=z,
\end{equation}
The solution can be obtained by doing a Fourier-Laplace transformation\cite{white_theory_2006}
\begin{equation}
  \chi=-\frac{z}{2}\int_{0}^{1}d\mu e^{-z^2\mu/2} (1-\mu^2)^{-1/4},
\end{equation}
and the integral required for matching condition is
\begin{equation}
\label{eq:matching}
  \int_{-\infty}^{0}\frac{dz}{z}\chi''+\int_{0}^{\infty}\frac{dz}{z}\chi''=\frac{2\pi \Gamma(3/4)}{\Gamma(1/4)}.
\end{equation}

Putting this result into Eqs. (\ref{eq:inner-matching}) and (\ref{eq:outer-matching}), we obtain the result of the integral in the region inside the resistive layer but outside the sublayer,
\begin{equation}
\label{eq:integral-layer}
\frac{1}{\psi_0}\left(\int_{-\delta_1}^{-\delta_2} \psi'' dx+\int_{\delta_2}^{\delta_1} \psi'' dx\right)=\frac{\gamma^{5/4}}{\eta^{3/4}k_c^{1/2}}\frac{2\pi \Gamma(3/4)}{\Gamma(1/4)}-\left(\frac{m}{k_{c}r_{s}}\right)^{2}2J'_{0}J'_{RE0}\delta_1 \ln\delta_1,
\end{equation}
where the higher order terms are ignored. The last term is calculated by substituting the real part of $\psi$ near $x=0$ into Eq. (\ref{eq:xipp}). Note that the last term is small compared to $\Delta'$ and can be ignored here. The matching condition that determines the growth rate is
\begin{equation}
\label{eq:gamma-condition}
\frac{\gamma^{5/4}}{\eta^{3/4}k_c^{1/2}}\frac{2\pi \Gamma(3/4)}{\Gamma(1/4)}=\Delta'-i\pi \frac{m J'_{RE0}}{|k_c| r_s},
\end{equation}
where we assume $\mathrm{Re}(\gamma)>0$. Comparing this result with Ref. \cite{furth_tearing_1973}, we find that the effect of runaway electron current on resistive tearing mode can be treated by adding an imaginary term to $\Delta'$. This correction term is of the same order of the original integral of $\psi''$ in the resistive layer. This will make $\gamma$ a complex number. The real part is the mode growth rate and the imaginary part means that the tearing mode will have a real frequency and propagate in the plasma frame. The amplitude of this imaginary term is proportional to $J_{RE0}'$ at the rational surface, thus for the case with $J_{RE0}'=0$ (like the tearing mode in slab geometry with a symmetric current profile, or a flat runaway electron current profile across the rational surface), the correction is zero. This imaginary part does not have a strong dependence on the value of $v_A/c$, as long as $|\delta_2|\ll|\delta_1|$. The sign of this term is determined by the sign of $J_{RE0}'$ and the streaming direction of the runaway electrons.

%
%

\section{Resistive kink mode with $m=1$}
\label{sec:kink}

For a resistive mode with $m=1$, the equation in the inner region is the same as Eqs. (\ref{eq:psipp}-\ref{eq:xipp}). However, the solution will match that of the internal kink mode in the outer region, which is $\xi=\xi_0$ for $r<r_s$, and $\xi=0$ for $r>r_s$ for plasma $\beta$ close to zero. This is a different boundary condition from the $m>1$ tearing mode. In addition, the perturbed flux $\psi$ will also be zero in the outer region, which means that the constant $\psi$ approximation cannot be used to calculate the inner region solution.

In previous studies of the $m=1$ resistive kink mode without runaway electron current, by ignoring the last two terms in Eq. (\ref{eq:xipp}), an exact solution is found\cite{coppi_resistive_1976,white_theory_2006},
\begin{equation}
\label{eq:xi-kink}
  \xi=\frac{\xi_0}{\sqrt{2\pi}}\int_z^\infty e^{-z^2/2} dz,
\end{equation}
\begin{equation}
\label{eq:psi-kink}
  \psi=\frac{\xi_0}{\sqrt{2\pi}}\frac{k_c \eta^{1/2}}{\gamma^{1/2}}\left[e^{-z^2/2}-z\int_{z}^{\infty}e^{-\zeta^2/2}d\zeta\right],
\end{equation}
where
\begin{equation}
z=\frac{x}{\hat{\delta}_1},\qquad \hat{\delta}_1=\frac{\eta^{1/2}}{\gamma^{1/2}}.
\end{equation}

The outer region gives the matching condition as
\begin{equation}
  \psi'(x=0+)=0,\qquad \psi'(x=0-)=\xi_0 k_c.
\end{equation}
Thus the growth rate satisfies
\begin{equation}
\label{eq:gamma-kink}
  \frac{\gamma^3}{\eta k_c}=\frac{\Delta \psi'}{\xi_0}=k_c.
\end{equation}

By including the runaway electron current term, as in Sec. \ref{sec:tearing}, the runaway electron current will cancel the last two terms in Eq. (\ref{eq:xipp}), making the growth rate close to the value given in Eq. (\ref{eq:gamma-kink}) in both the small and large $\eta$ cases. Like in Sec. \ref{sec:tearing}, we find that $j_{RE}$ term can cancel the last two terms in Eq. (\ref{eq:xipp}), leaving the combined equation of $\xi$ exactly the same as the approximate equation in \cite{coppi_resistive_1976}. The solution of $\xi$ will also be the same as in Eq. (\ref{eq:xi-kink}). The form of $\psi$, however, will be affected by the presence of the last two terms in Eq. (\ref{eq:xipp}). Here we use the form in Eq. (\ref{eq:psi-kink}) as the leading order solution of $\psi$, together with Eq. (\ref{eq:xipp}) to calculate the next order solution of $\psi''$.

Like the derivation in Sec. \ref{sec:tearing}, the introduction of runaway electron current can give rise to a sublayer of $\psi$ near $x=0$ with the width $|\delta_2|$ in Eq. (\ref{eq:delta-2}). This leads to an additional imaginary term in the integral of $\psi''$ across the sublayer. Using the result of $\psi(x=0)$ calculated from Eq. (\ref{eq:psi-kink}), we get
\begin{equation}
\int_{-\delta_2}^{\delta_2} \psi'' dx\approx i\sqrt{\frac{\pi}{2}}\frac{ \eta^{1/2}}{\gamma^{1/2}} \frac{J'_{RE0}}{ r_s} \xi_0.
\end{equation}

In addition, the second term in Eq. (\ref{eq:psi-kink}) can also contribute to the integral of $\psi''$ in the region inside the resistive layer but outside the sublayer, through the last two terms on the right side of Eq. (\ref{eq:xipp}),
\begin{equation}
\label{eq:integral-layer2}
\int_{-\hat{\delta}_1}^{-\delta_2} \frac{a\psi}{x} dx+\int_{\delta_2}^{\hat{\delta}_1} \frac{a\psi}{x} dx=-\frac{J'_{0}}{ r_s}\frac{\xi_{0}\hat{\delta}_{1}}{\sqrt{2\pi}}\int_{-1}^{1}dz\int_{z}^{\infty}e^{-\zeta^{2}/2}d\zeta=-\frac{J'_{0}}{ r_s}\xi_{0}\hat{\delta}_{1}.
\end{equation}
In addition, there is also  a contribution from the real part of $\psi$ near $x=0$ like the last term in Eq. (\ref{eq:integral-layer}). However, calculation shows that this term is proportional to $\hat{\delta}_1^2 \ln\hat{\delta}_1$, which is  a higher order effect compared to Eq. (\ref{eq:integral-layer2}). 

Combining these two additional terms with the integral of $\xi''$, we obtain the new matching condition
\begin{equation}
\label{eq:gamma-condition-kink}
  \frac{1}{\xi_0}\left(\int_{-\hat{\delta}_1}^{-\delta_2} \psi'' dx+\int_{\delta_2}^{\hat{\delta}_1} \psi'' dx\right)=\frac{\gamma^3}{\eta k_c}=k_c-\frac{ \eta^{1/2}}{\gamma^{1/2}}\frac{1}{r_s}\left(i\sqrt{\frac{\pi}{2}}J'_{RE0}-J'_{0}\right).
\end{equation}
Note that the last term in Eq. (\ref{eq:gamma-condition-kink}) does not depends on $J_{RE}$, which is simply a correction term due to the last two term in Eq. (\ref{eq:xipp}), and will lead to a change in the growth rate rather than real frequency. Given $\eta^{1/2}/\gamma^{1/2}\sim \eta^{1/3} \ll 1$, the correction term coming from runaway electron current is of higher order compared to the original solution, which is different from the results in Eq. (\ref{eq:gamma-condition}). However, for plasma with large resistivity like in a post-disruption scenario discussed in Sec. \ref{sec:simulation}, the correction can still be important. In addition, the runaway electron current will also contribute to the real part of the dispersion relation, which will change the value of the growth rate and make it deviate from the 1/3 power law for large $\eta$ cases. 

\section{Compare with numerical result and M3D-C$^1$ simulation result}
\label{sec:simulation}

In this section we compare the MHD instability growth rates and real frequencies obtained in Secs. \ref{sec:tearing} and \ref{sec:kink} with both an eigenvalue code and M3D-C$^1$ simulation result. The eigenvalue code solves Eqs. (\ref{eq:gammapsi}-\ref{eq:jre}) using the finite difference method. For the M3D-C$^1$ code, we use the 2-field linear version, which only solves for $\phi$ for the velocity field and $\psi$ for the magnetic field, in order to benchmark with reduced MHD results. The MHD equations and runaway electron equations are solved on a 2D circular plane, and the toroidal derivative terms are represented with a spectral method. To resolve the runaway current sublayer structure, fine mesh packing is applied near the rational surface where the mode is excited. The results of linear simulations with full MHD equations are discussed in Ref. \cite{zhao_simulation_2020}.

Before proceeding with numerical results, it is useful to make some simple estimates on the value of normalized resistivity $\eta$, on which the growth rates and the real frequencies will depend. For a runaway electron experiments in quiescent operation in DIII-D, where ions are all deuterium and $B=2$T, electron density $n_e=10^{21}$m$^{-3}$, electron temperature $T_e=1$ keV, $a=0.67$m, the normalized resistivity $\eta$ is only about $2\times 10^{-9}$, and $c/v_A=30$. However, for a post-disruption scenario in ITER, with $B=5$T, $n_e=10^{21}$m$^{-3}$, $T_e=5$ eV, $a=2$m, $\eta\approx 10^{-5}$ and $c/v_A\approx 123$. In addition, if the plasma is composed by argon instead of deuterium due to impurity injection, the values of $\eta$ and $c/v_A$ can increase to $4\times 10^{-5}$ and 550.

\begin{figure}[ht]
	\begin{center}
		\includegraphics[width=0.45\linewidth]{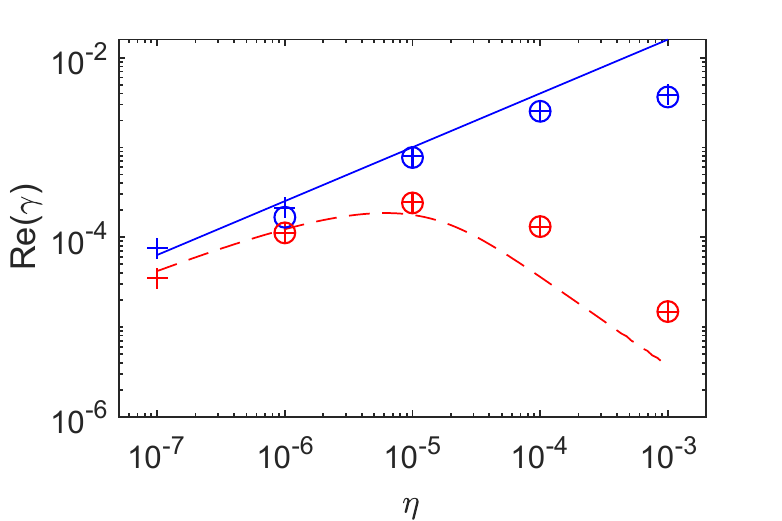}
		\includegraphics[width=0.45\linewidth]{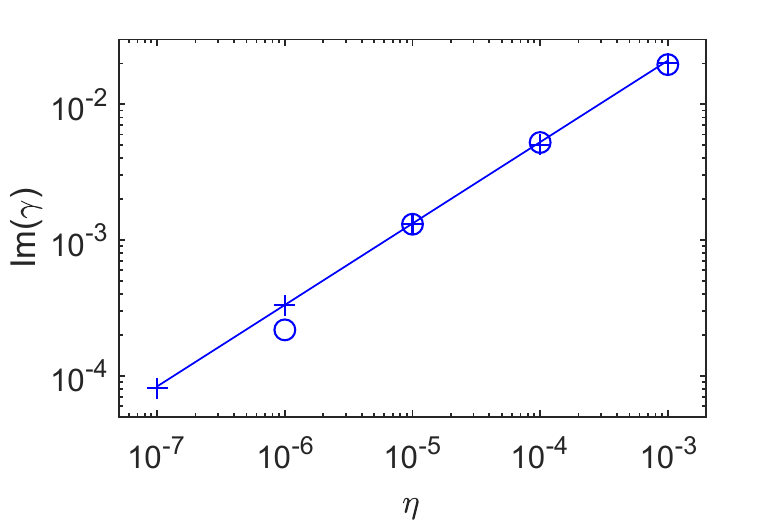}
	\end{center}
	\caption{\label{fig:tearing} Real and imaginary parts of $\gamma$ for (2,1) tearing mode. Blue lines and points characterize the results with runaway electron current, and red lines and points represent the results without. Solid lines are values calculated from Eq. (\ref{eq:gamma-condition}). Dashed lines are values calculated from Ref. \cite{militello_effects_2004}. Plus sign symbols (+) show the result from eigenvalue code. Circle symbols ($\circ$) show the result from M3D-C$^1$ simulation.}
\end{figure}

We first compare the results of $m>1$ resistive tearing modes from these different methods. The MHD equilibrium in a cylinder is set up with $q=1.15(1+r^2/0.6561)$. For the case with runaway electron current, all equilibrium current is assumed to be carried by runaway electrons ($J_0=J_{RE0}$). In Fig. \ref{fig:tearing} we show the values of $\gamma$ for a $m=2,n=1$ mode, for cases both with runaway electron current and without runaway electron current. We can see that in the case without runaway electron current, the growth rate from both the eigenvalue code and the M3D-C$^1$ simulation deviate from the 3/5 power law\cite{furth_tearing_1973} for large $\eta$, but agree with the result in Ref. \cite{militello_effects_2004}. After including the runaway electron term in the equation, the growth rates from both numerical codes agree well with Eq. (\ref{eq:gamma-condition}) except for very large $\eta$. This is consistent with the numerical calculation in Ref. \cite{helander_resistive_2007}. The real frequency of the modes from both the numerical codes also agree well with Eq. (\ref{eq:gamma-condition}), and are of the same order as the growth rate for all values of $\eta$. Since in all cases $|\delta_2|\ll|\delta_1|$, it is found that both the growth rates and real frequencies are insensitive to the value of $c/v_{A}$, consistent with the results reported in Ref. \cite{zhao_simulation_2020}. The deviation at very large $\eta$ is due to the fact that the large value of the tearing layer width $\delta$  can become comparable to $a$ in these cases, which is the non-asymptotic regime. Thus the assumption we took in the derivations in Sec. \ref{sec:tearing}, such as constant-$\psi$ approximation and $x/r_s\ll 1$, are no longer valid.

\begin{figure}[ht]
	\begin{center}
		\includegraphics[width=0.45\linewidth]{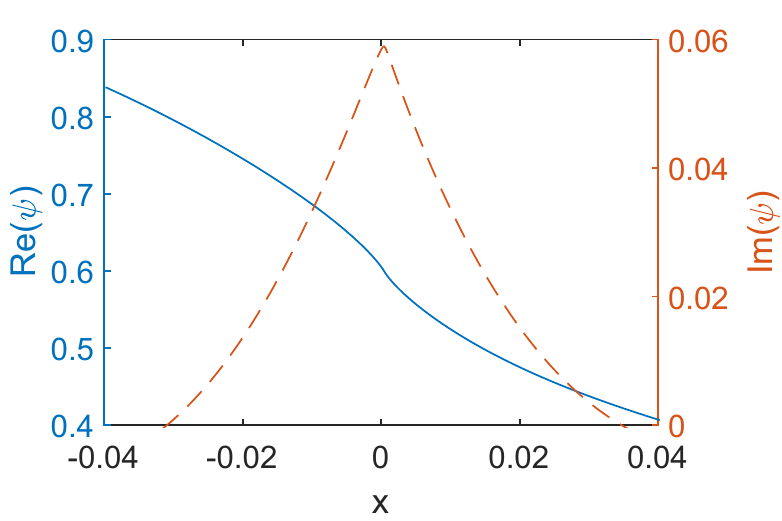}
		\hspace{1em}
		\includegraphics[width=0.45\linewidth]{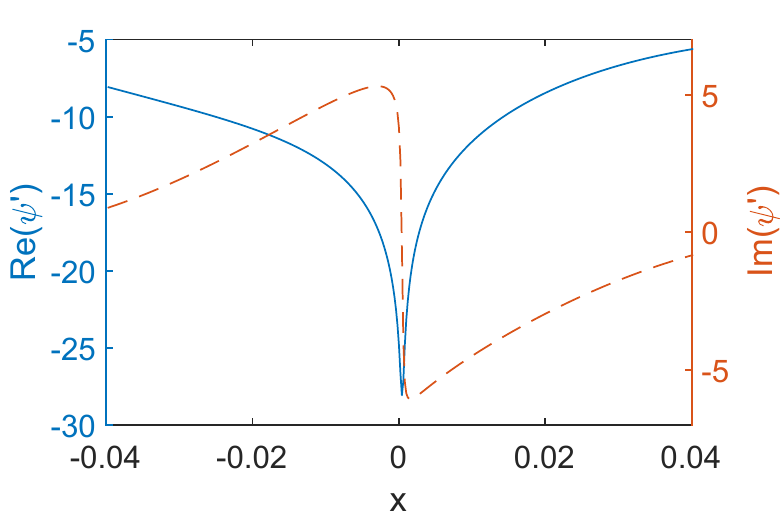}
	\end{center}
	\caption{\label{fig:psi} Values of $\psi$ and $\psi'$ near rational surface from eigenvalue code result for (2,1) mode, including both real (solid) and imaginary parts (dashed).}
\end{figure}

Fig. \ref{fig:psi} shows the real and imaginary parts of $\psi$ and $\psi'$ for the (2,1) modes near the rational surface, from the eigenvalue code, for $\eta=10^{-5}$ and $c/v_{A}=40$. We can see clearly the existence of a sublayer structure. Due to the existence of large $j_{RE}$ near $x=0$, the imaginary part of $\psi'$ will behave like a step function. The change of $\psi'$ inside the sublayer will then be compensated at $|\delta_2|<|x|<|\delta_1|$ to satisfy the boundary condition at the outer bounds. The real part of $\psi'$ is also affected by $j_{RE}$ and behaves like $\sim\ln |x|$ near $x=0$, as discussed in Sec. \ref{sec:tearing}.

\begin{figure}[ht]
	\begin{center}
		\includegraphics[width=0.45\linewidth]{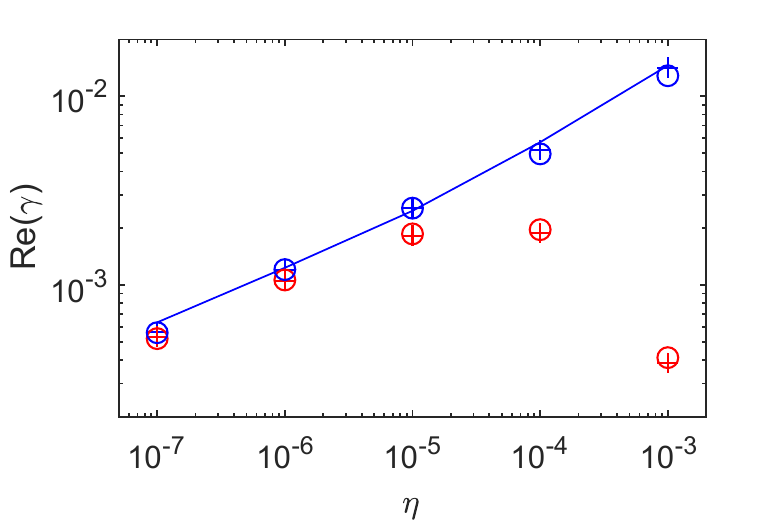}
		\includegraphics[width=0.45\linewidth]{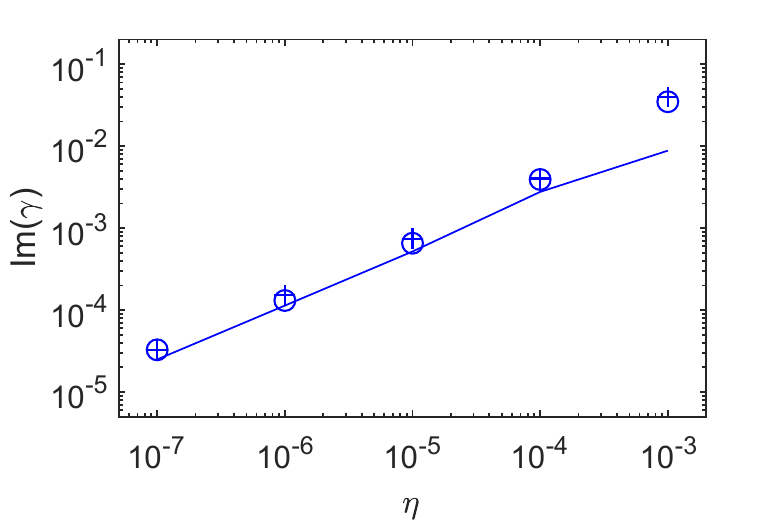}
	\end{center}
	\caption{\label{fig:kink} Real and imaginary parts of $\gamma$ for (1,1) kink mode, with (blue) and without (red) runaway electron current. Solid lines are values calculated from Eq. (\ref{eq:gamma-condition-kink}). Plus sign symbols (+) show the result from eigenvalue code.  Circle symbols ($\circ$) show the result from M3D-C$^1$ simulation.}
\end{figure}

We next compare $\gamma$ of (1,1) resistive kink modes. The safety factor profile is now set to be $q=0.9(1+x^2/2)$ to incorporate a $q=1$ surface. The results for the growth rates and real frequencies are shown in Fig. \ref{fig:kink}. The analytical results are calculated from Eq. (\ref{eq:gamma-condition-kink}) using an iterative method. Similar to those for (2,1) modes, the growth rates deviate from the 1/3 power law for large $\eta$ without runaway electron current, but with runaway electron current the results follow the 1/3 power law except for very large $\eta$. This is due to the correction to the real part in Eq. (\ref{eq:gamma-condition-kink}). These results are consistent with those in Ref. \cite{matsuyama_reduced_2017}. The real frequencies are much smaller than the growth rate for small values of $\eta$, but for large $\eta$ the real frequency can be even larger than the growth rate. The results for very large $\eta$ deviate from Eq. (\ref{eq:gamma-kink}) because the very large layer width breaks down the assumption used in the calculations in Sec. \ref{sec:kink}.

\section{Conclusion}
\label{sec:conclusion}

In this paper we establish a new theoretical model to understand the linear properties of both resistive tearing and resistive kink modes in the presence of runaway electron current. The model is based on the asymptotic analysis method for the reduced MHD equations in cylindrical geometry as presented in previous studies of runaway electron current’s effect\cite{helander_resistive_2007}. Extending from the analysis in Ref. \cite{helander_resistive_2007}, we studied the sublayer structure of the mode formed by the convection equation of runaway electron current, and its effect on the mode dispersion relation. It is found that the sublayer can have a strongly peaked perturbed current, whose width can be much smaller than the resistive layer and depends on the value of $v_{A}/c$. On the other hand, this runaway electron sublayer can give rise to an imaginary term in the dispersion relation whose amplitude does not depend on the value of $v_{A}/c$, which is similar to the pole contribution in a Landau contour. The growth rates and real frequencies of modes calculated using this dispersion relation are compared with results obtained using an eigenvalue calculation and M3D-C$^{1}$ simulations.

The real frequency of the mode found in the paper is proportional to the gradient of runaway electron current $J_{RE}'$ at the rational surface. This current gradient will break the symmetry of the profile across the rational surface, making an odd part in the perturbed flux function. As shown in Ref. \cite{militello_effects_2004} and in Eq. \ref{eq:gamma-condition-kink}, this gradient can also change the mode growth rate. In addition, as discussed in Ref. \cite{hastie_nonlinear_2005}, the break of symmetry can affect the saturation level of tearing mode. It is believed that such effect can be important for disruptive plasma since runaway electron current can have a more peaked profile than thermal plasma\cite{eriksson_current_2004}.
  
With the simplified model presented here, including a mode resonant at a single surface and a perfectly conducting wall, the effect of the runaways on the layer solution, frequency and growth rate are clearly important.  Given multiple rational surfaces and/or a surrounding resistive wall and bulk plasma flow, the contributions from the layer response at each surface, and the currents driven in the wall, will again combine to select a single frequency of the mode.  The mode frequency determines the rate of propagation in the plasma frame and, via a Doppler shift, relative to the resistive wall.  The stability and structure of the mode will then depend on the conditions at all the surfaces, the plasma flow, and the properties of the resistive wall.  Given a significant runaway current and gradient across surfaces, ultimately the stability of the runaway electron confining plasma depends on this real frequency, and thus so does our ability to keep the runaways confined. Clearly, including this physics into MHD stability calculations of experiments with runaway electron generation is critically important.

The result of this work can help future numerical and theoretical simulation research on runaway electrons. The very thin layer of perturbed current inside the resistive layer poses a significant challenge to simulation of MHD modes with runaway electrons, especially for $m>1$ tearing modes. To resolve the mode structure and get the correct value of mode real frequency, either a fine mesh packing is required near the rational surface, or one has to use a reduced unphysical value of $c/v_{A}$ like in Ref.\cite{cai_influence_2015}. In addition, the width of this sublayer can be comparable to the electron skin depth ($d_e=c/\omega_{pe}$, where $\omega_{pe}$ is the plasma oscillation frequency). This means that inside the sublayer, the electron inertial term ($d_e^2 dj/dt)$ can become important and should be included in the Ohm's law. By including this term, we find that for ITER disruption case ($\eta\sim 10^{-5}$), the inertial term is still a subdominant term compared to the resistive term ($\eta j$). But for case with smaller $\eta$ like the quiescent runaway electron experiments, the inertial term can be important. The inertial term has been shown to play a role in the fast Hamiltonian reconnection\cite{ottaviani_nonlinear_1993,grasso_hamiltonian_1999,bhattacharjee_current_2005}, and give rise to a singular current spike. The nonlinear evolution of the modes can also be strongly affected by the existence of this sublayer and current spike, the study of which we leave for future work.

\begin{acknowledgments}
Chang Liu would like to thank Per Helander, Guo-Yong Fu, Hank R. Strauss and Akinobu Matsuyama for fruitful discussion. This work was supported by the Simulation Center of electrons (SCREAM) SciDAC center by Office of Fusion Energy Science and Office of Advanced Scientific Computing of U. S. Department of Energy, under contract No. DE-SC0016268 and DE-AC02-09CH11466. This research used the high-performance computing cluster at Princeton Plasma Physics Laboratory and the Eddy cluster at Princeton University. The data that support the findings of this study are available from the corresponding author upon reasonable request.
\end{acknowledgments}

\section*{DATA AVAILABILITY}
The data that support the findings of this study are available from the corresponding author upon reasonable request.

\bibliography{paper}

\end{document}